\documentclass{PoS}
\usepackage{graphicx}
\usepackage[dvips]{epsfig}
\usepackage{amssymb,amsmath}
\usepackage{slashed}
\usepackage{url}

\title{Probing neutrino parameters with \\ a Two-Baseline Beta-beam set-up}  

\ShortTitle{Two-Baseline Beta-beam set-up}

 \author{\speaker{Sanjib Kumar Agarwalla}\thanks{It is my pleasure 
  to acknowledge the support of Harish-Chandra Research Institute (HRI) 
  where the work has been done. The computational work presented in this 
  talk has been performed using the HRI cluster facilities. 
  I would like to thank Anselmo Cervera for providing me the local 
  hospitality during Nufact08.} \\
         Department of Physics, Virginia Tech, \\
	 Blacksburg, VA 24061, USA \\
         E-mail: \email{sanjib@vt.edu}}

 \author{Sandhya Choubey  \\
         Harish-Chandra Research Institute, \\
	 Chhatnag Road, Jhunsi, Allahabad  211019, India \\
         E-mail: \email{sandhya@hri.res.in}}

 \author{Amitava Raychaudhuri  \\
         Harish-Chandra Research Institute, \\
	 Chhatnag Road, Jhunsi, Allahabad  211019, India \\
         E-mail: \email{raychaud@hri.res.in}}

\abstract{
We discuss the prospects of exploring the neutrino mass parameters
with a CERN based Beta-beam experiment using two different detectors
at two different baselines. The proposed set-up consists of a 50 kton iron
calorimeter (ICAL) at a baseline of around 7150 km which is roughly the
magic baseline, e.g., ICAL@INO, and a 50 kton Totally Active Scintillator Detector
at a distance of 730 km, e.g., at Gran Sasso. We take $^8$B and $^8$Li 
source ions with a boost factor $\gamma$ of 650 for the magic baseline while
for the closer detector we consider $^{18}$Ne and $^6$He ions with a
range of Lorentz boosts. We find that the locations of the two
detectors complement each other leading to an exceptional high
sensitivity. With $\gamma=650$ for $^8$B/$^8$Li and $\gamma=575$ for
$^{18}$Ne/$^6$He and total luminosity corresponding to 
$5\times (1.1\times 10^{19})$ and $5\times (2.9\times 10^{19})$ 
useful ion decays in neutrino and antineutrino modes respectively, we find 
that the two-detector set-up can probe maximal CP violation and establish
the neutrino mass ordering if $\sin^22\theta_{13}$ is $1.8 \times 10^{-5}$ 
and $4.6 \times 10^{-5}$, respectively, or more.
The sensitivity reach for  $\sin^22\theta_{13}$ itself is $5.3 \times 10^{-5}$.
CP violation can be discovered for 64\% of the possible $\delta_{CP}$ values for
$\sin^22\theta_{13} \geq 8\times 10^{-5}$.}

\FullConference{10th International Workshop on Neutrino Factories, Super beams and Beta beams\\
		 June 30 - July 5 2008\\
		 Valencia, Spain}

\newcommand{\ma}{\Delta m^2_{31}}
\newcommand{\stch}{\sin^2 2\theta_{13}}
\newcommand{\stcht}{\sin^2 2\theta_{13}{\mbox {(true)}}}
\newcommand{\dcpt}{\delta_{CP}{\mbox {(true)}}}
\newcommand{\sig}{$3\sigma$}

\begin{document}

We consider a two-baseline Beta-beam \cite{twobaseline} set-up,
one with $L=7152$ km, the CERN-INO baseline \cite{betaino}, and
another with $L=730$ km which is the CERN-Gran Sasso (LNGS) distance.
For the CERN-INO case $^8$B and $^8$Li are the preferred source ions
and we take $\gamma=650$. For the CERN-LNGS set-up, on the other hand,
we choose the $^{18}$Ne and $^6$He ions and allow their $\gamma$ to
vary between 250-650. Since the $^8$B and $^8$Li ions would produce
multi-GeV neutrino beams for $\gamma=650$, we use a 50 kton iron
calorimeter (ICAL) for the longer baseline at INO \cite{ino}.
For the intermediate baseline option, since we are
interested in the lower energy $^{18}$Ne and $^6$He ions,
we assume a 50 kton Totally Active Scintillator Detector (TASD) in order to
harness the low energy events required for better CP sensitivity.
We present results for $5\times (1.1 \times 10^{19})$
and $5\times (2.9\times 10^{19})$ useful ion decays
in neutrino and antineutrino modes respectively for both
baseline set-ups. We also show the projected sensitivity
with one order less statistics.

\begin{figure}[!t]
\includegraphics[width=7.25cm,height=.28\textheight]{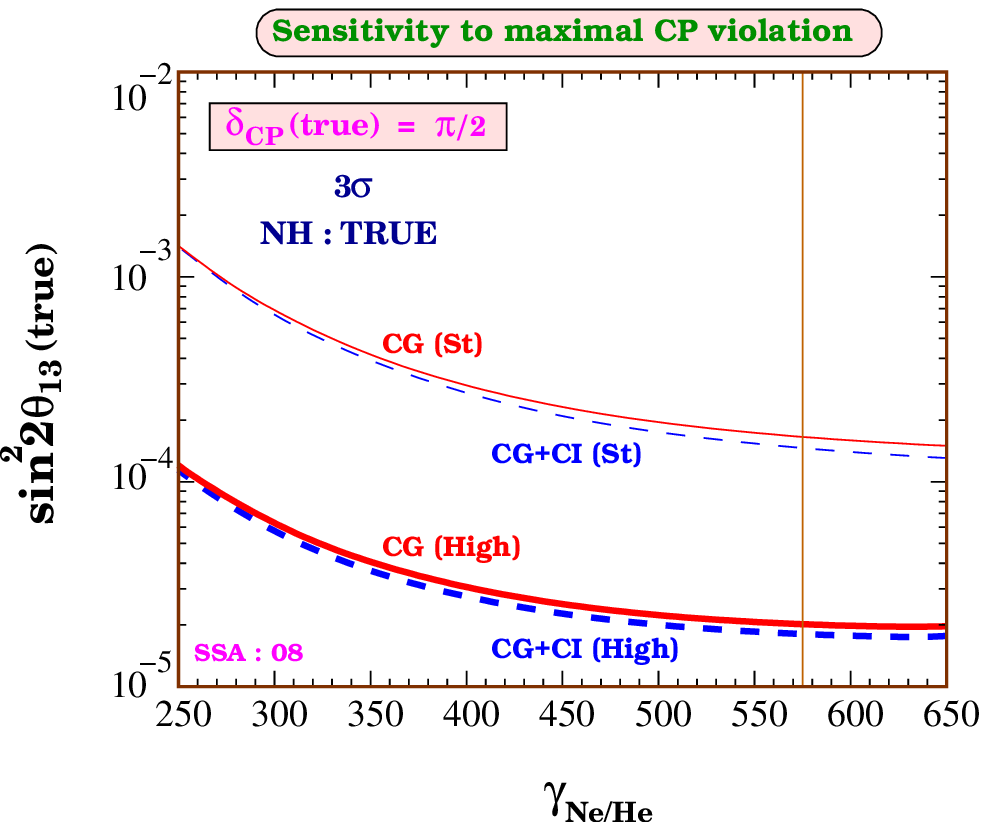}
\hglue 0.5cm
\includegraphics[width=7.25cm,height=.28\textheight]{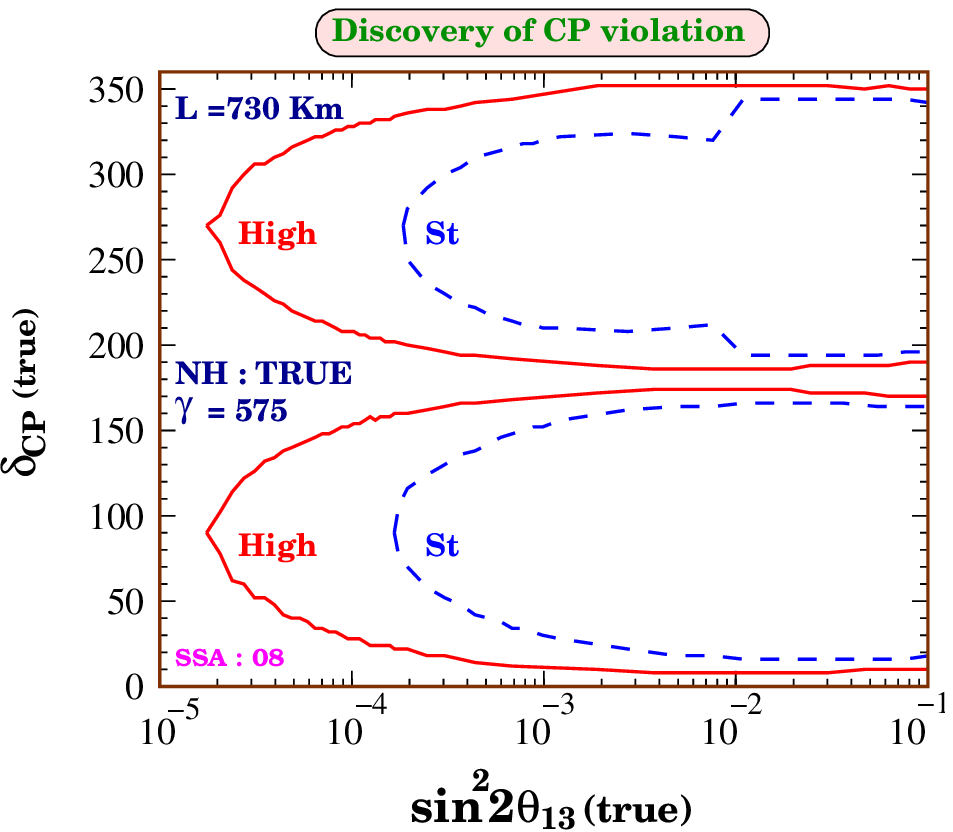}
\caption{\label{fig:cp}
Left panel shows the $3\sigma$ $\stcht$ reach for sensitivity to
``maximal CP violation''. The red solid curves (marked as `CG')
are for the CERN-TASD@LNGS alone while the blue dashed lines
(marked as `CG+CI') are for the combined data from CERN-TASD@LNGS
and CERN-ICAL@INO. The results are shown as a function of the Lorentz
boost for $^{18}$Ne and $^6$He (taken same for both ions), for
$\dcpt=90^\circ$. Thick lines (marked ``High") are for $5\times  (1.1 \times
10^{19})$ useful $^{18}$Ne and $^8$B decays and $5\times (2.9\times
10^{19})$ useful $^6$He and $^8$Li decays, while thin lines (marked ``St")
are for $5\times  (1.1 \times 10^{18})$ and $5\times (2.9\times 10^{18})$
useful ion decays respectively. In the right panel, the area enclosed by the
curves represents the $3\sigma$ range of $\dcpt$ as a function of $\stcht$
for which the data can be used to rule out the CP-conserving scenario
using the CERN-Gran Sasso reference TASD set-up with $^{18}$Ne and
$^6$He as source ions.
}
\end{figure}

\begin{figure}[!t]
\includegraphics[width=7.25cm,height=.28\textheight]{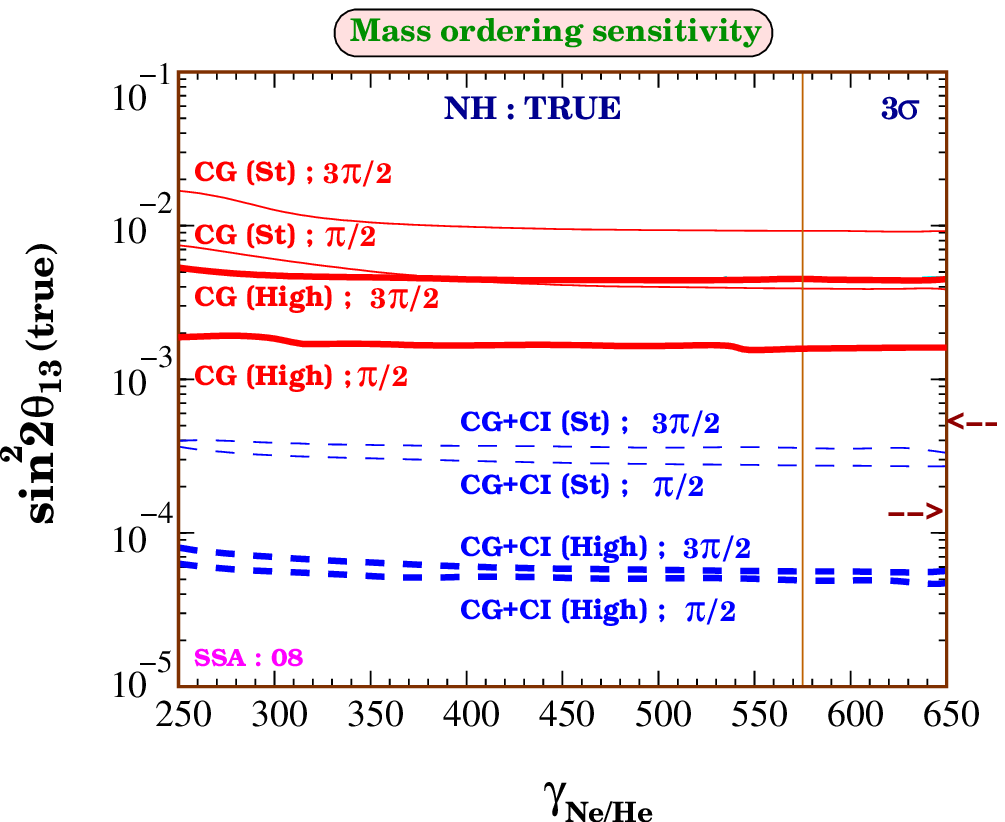}
\hglue 0.5cm
\includegraphics[width=7.25cm,height=.28\textheight]{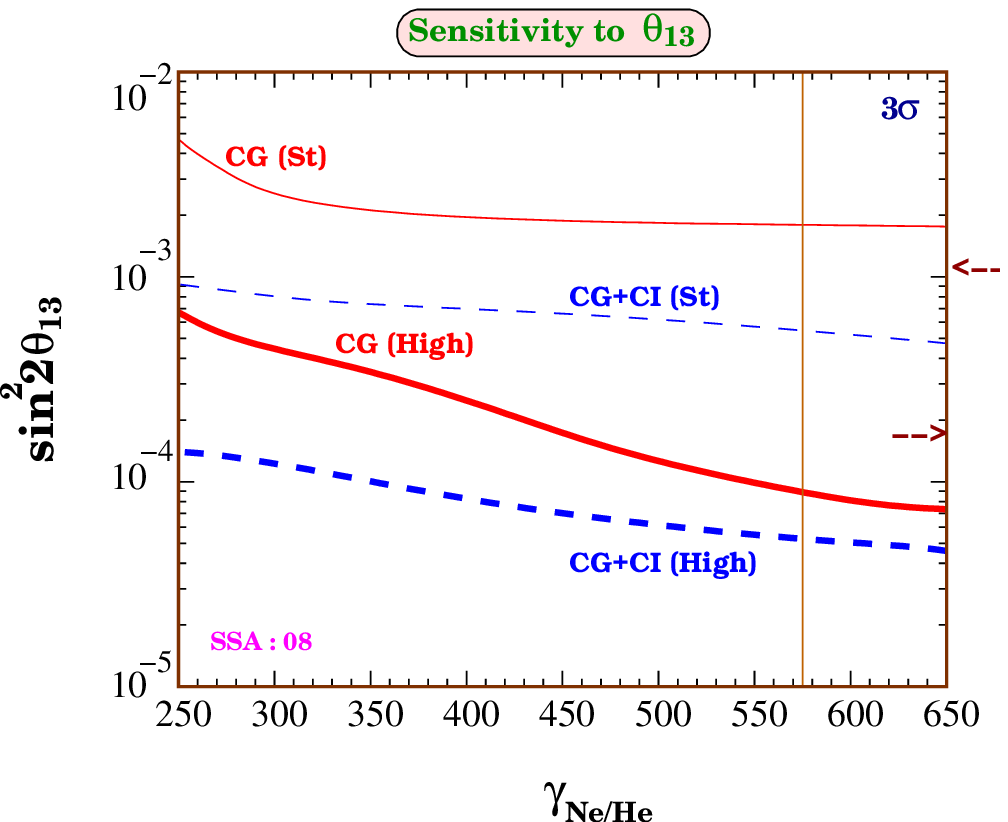}
\caption{\label{fig:senshier}
The left and right panels depict the $sgn(\ma)$ sensitivity
reach and the $\stch$ sensitivity reach, respectively, at
$3\sigma$ as a function of the boost factor for $^{18}$Ne and $^6$He.
In both panels, the red solid lines are for CERN-TASD@LNGS alone
while the blue dashed lines are for the combined data from
CERN-ICAL@INO and CERN-TASD@LNGS. Results for $\dcpt = \pi/2$ and
$3\pi/2$ are shown.  Thick lines are for $5\times (1.1 \times
10^{19})$ useful $^{18}$Ne and $^8$B decays and $5\times (2.9\times
10^{19})$ useful $^6$He and $^8$Li decays, while thin lines are for
$5\times (1.1 \times 10^{18})$ and $5\times (2.9\times 10^{18})$
useful ion decays respectively. The sensitivity reaches for the
CERN-ICAL@INO set-up alone are indicated for both luminosities
(for $\dcpt$ = 0 in the left panel) by arrows on the right side of the panels.
The location of $\gamma$ = 575 is shown.
}
\end{figure}

\begin{table}[!t]
\begin{center}
{\footnotesize
\begin{tabular}{|c||c|c||c|c||c|c|} \hline\hline
Set-up
& \multicolumn{2}{|c||}{\rule[0mm]{0mm}{2mm}$\stch$ Discovery (\sig)}
& \multicolumn{2}{|c||}{\rule[0mm]{0mm}{2mm}Mass Ordering (\sig)}
& \multicolumn{2}{|c|}{\rule[0mm]{0mm}{2mm}Maximal CP violation (\sig)}
\cr \cline{2-7}
\hline\hline
CERN-INO
& \multicolumn{2}{|c||}{{\rule[0mm]{0mm}{0mm}}}
& \multicolumn{2}{|c||}{\rule[0mm]{0mm}{0mm}{}}
& \multicolumn{2}{|c|}{\rule[0mm]{0mm}{0mm}{}}
\cr
$^8$B+$^8$Li, $\gamma = 650$ & \multicolumn{2}{|c||}
{{\rule[0mm]{0mm}{0mm}
$9.5\times 10^{-5}$
}}
& \multicolumn{2}{|c||}{{\rule[0mm]{0mm}{0mm}
$9.4\times 10^{-5}$
}}
& \multicolumn{2}{|c|}{{\rule[0mm]{0mm}{0mm}
Not possible
}}
\cr
\hline\hline
CERN-LNGS
& \multicolumn{2}{|c||}{{\rule[0mm]{0mm}{0mm}}}
& \multicolumn{2}{|c||}{{\rule[0mm]{0mm}{0mm}}}
& \multicolumn{2}{|c|}{{\rule[0mm]{0mm}{0mm}}}
\cr
$^{18}$Ne+$^6$He, $\gamma = 575$ & \multicolumn{2}{|c||}{{\rule[0mm]{0mm}{0mm}
$2.07\times 10^{-5}$
}}
& \multicolumn{2}{|c||}{{\rule[0mm]{0mm}{0mm}
$1.58\times 10^{-3}$
}}
& \multicolumn{2}{|c|}{{\rule[0mm]{0mm}{0mm}
$1.97\times 10^{-5}$
}}
\cr
\hline\hline
CERN-LNGS
& \multicolumn{2}{|c||}{{\rule[0mm]{0mm}{0mm}}}
& \multicolumn{2}{|c||}{\rule[0mm]{0mm}{0mm}{}}
& \multicolumn{2}{|c|}{\rule[0mm]{0mm}{0mm}{}}
\cr
$^{18}$Ne+$^6$He, $\gamma = 575$
& \multicolumn{2}{|c||}{{\rule[0mm]{0mm}{0mm}}}
& \multicolumn{2}{|c||}{{\rule[0mm]{0mm}{0mm}}}
& \multicolumn{2}{|c|}{{\rule[0mm]{0mm}{0mm}}}
\cr
$+$ &
\multicolumn{2}{|c||}{{\rule[0mm]{0mm
}{0mm}
$1.88\times 10^{-5}$
}}
& \multicolumn{2}{|c||}{{\rule[0mm]{0mm}{0mm}
$4.64\times 10^{-5}$
}}
& \multicolumn{2}{|c|}{{\rule[0mm]{0mm}{0mm}
$1.78\times 10^{-5}$
}}
\cr
CERN-INO
& \multicolumn{2}{|c||}{{\rule[0mm]{0mm}{0mm}}}
& \multicolumn{2}{|c||}{\rule[0mm]{0mm}{0mm}{}}
& \multicolumn{2}{|c|}{\rule[0mm]{0mm}{0mm}{}}
\cr
$^8$B+$^8$Li, $\gamma = 650$
& \multicolumn{2}{|c||}{{\rule[0mm]{0mm}{0mm}}}
& \multicolumn{2}{|c||}{\rule[0mm]{0mm}{0mm}{}}
& \multicolumn{2}{|c|}{\rule[0mm]{0mm}{0mm}{}}
\cr
\hline\hline
Optimized
& \multicolumn{2}{|c||}{{\rule[0mm]{0mm}{0mm}}}
& \multicolumn{2}{|c||}{\rule[0mm]{0mm}{0mm}{}}
& \multicolumn{2}{|c|}{\rule[0mm]{0mm}{0mm}{}}
\cr
Neutrino Factory &
\multicolumn{2}{|c||}{{\rule[0mm]{0mm}{0mm}
$1.5\times 10^{-5}$
}}
& \multicolumn{2}{|c||}{{\rule[0mm]{0mm}{0mm}
$1.5\times 10^{-5}$
}}
& \multicolumn{2}{|c|}{{\rule[0mm]{0mm}{0mm}
$1.5\times 10^{-5}$
}}
\cr
\hline\hline

\end{tabular}
\caption{\label{tab:compare}
Results are shown for a five-year run with the luminosity
$1.1 \times 10^{19}$ ($2.9\times 10^{19}$) useful ion decays 
per year in the $\nu$ ($\bar\nu$) mode. The numbers correspond 
to $\dcpt=90^\circ$. For comparison, the expectations from an optimized
two-baseline Neutrino Factory set-up with upgraded magnetized iron detectors 
are also listed \cite{issphysics,nufactoptim}.}
}
\end{center}
\end{table}

The $3\sigma$ sensitivity to ``maximal CP violation'' is
presented in Fig. \ref{fig:cp}. The two-baseline
combined results for $sgn(\ma)$ and $\sin^22\theta_{13}$
sensitivity reach have been depicted in Fig. \ref{fig:senshier}.

In conclusion, we present Table \ref{tab:compare} containing
all the essential results.


\end{document}